\documentclass[conference]{IEEEtran}
\IEEEoverridecommandlockouts
\usepackage{cite}
\usepackage{amsmath,amssymb,amsfonts}
\usepackage{algorithmic}
\usepackage{graphicx}
\usepackage{textcomp}
\usepackage{xcolor}
\usepackage{hyperref}
\usepackage{censor}
\usepackage{subcaption}
\usepackage{booktabs}
\usepackage{tabularx}

\usepackage[a4paper, total={184mm,239mm}]{geometry}
\def\BibTeX{{\rm B\kern-.05em{\sc i\kern-.025em b}\kern-.08em
    T\kern-.1667em\lower.7ex\hbox{E}\kern-.125emX}}

\newcommand{\implication}[1]{%
    \begin{center}
        \fbox{%
            \begin{minipage}{0.95\linewidth}
                \textbf{Implication}: #1
            \end{minipage}%
        }
    \end{center}%
}

\begin{document}

\title{Towards Multi-dimensional Elasticity for\\Pervasive Stream Processing Services}

%     \author{
% Anonymous Authors\\[0.2cm]

% \IEEEauthorblockA{Anonymous Affiliations\\   Anonymous E-Mail\\
%     }}

\author{
Boris Sedlak\IEEEauthorrefmark{1}, Andrea Morichetta\IEEEauthorrefmark{1}, Philipp Raith\IEEEauthorrefmark{1}, Víctor Casamayor Pujol\IEEEauthorrefmark{2}, and Schahram Dustdar\IEEEauthorrefmark{1}\IEEEauthorrefmark{2}\\

\IEEEauthorblockA{\IEEEauthorrefmark{1}\textit{Distributed Systems Group}, 
    Vienna University of Technology (\textit{TU Wien}), Vienna 1040, Austria.\\   Email: \{b.sedlak, a.morichetta, p.raith, dustdar\}@dsg.tuwien.ac.at
}

\IEEEauthorblockA{\IEEEauthorrefmark{2}\textit{Engineering Department}, 
Universitat Pompeu Fabra (\textit{UPF}), Barcelona 08018, Spain.
% \\ Email:\{victor.casamayor@upf.edu\}
}
}

\maketitle

\begin{abstract}
This paper proposes a hierarchical solution to scale streaming services across quality and resource dimensions. Modern scenarios, like smart cities, heavily rely on the continuous processing of IoT data to provide real-time services and meet application targets (Service Level Objectives -- SLOs). While the tendency is to process data at nearby Edge devices, this creates a bottleneck because resources can only be provisioned up to a limited capacity. To improve elasticity in Edge environments, we propose to scale services in multiple dimensions -- either resources or, alternatively, the service quality. We rely on a two-layer architecture where (1) local, service-specific agents ensure SLO fulfillment through multi-dimensional elasticity strategies; if no more resources can be allocated, (2) a higher-level agent optimizes global SLO fulfillment by swapping resources. The experimental results show promising outcomes, outperforming regular vertical autoscalers, when operating under tight resource constraints.
\end{abstract}

\begin{IEEEkeywords}
Elasticity, Edge Intelligence, Reinforcement Learning, Service Level Objectives, Stream Processing
\end{IEEEkeywords}

\section{Introduction}

% \ins{@Philipp: please revise S1 and verify the general utility of the approach, e.g., for serverless, when it can improve the utilization of devices because it must allocate fewer resources than in the agreement because it can also scale the quality.}

\begin{figure}[t]
    \centering
    \includegraphics[width=1.0\linewidth]{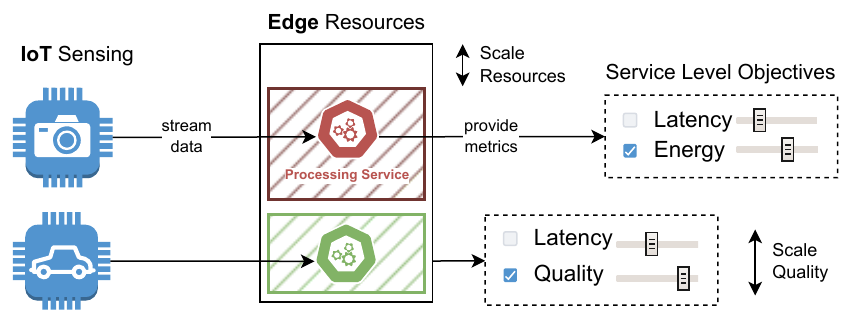}
    \caption{Processing IoT data at resource-constrained devices; if SLOs are violated, scale either resources or service quality}
    \label{fig:high-level-idea}
\vspace{-6pt}
\end{figure}

IoT adoption has grown significantly in areas where it provides clear value to human society, such as home automation, smart health, and smart city infrastructure~\cite{chui2021internet} -- these use cases are characterized by incessant amounts of streamed data. Smart city scenarios, for example, can involve collecting and analyzing video and images, sound, movement, and temperatures through stream processing services. On top of this, these services must operate within runtime requirements, e.g., latency or quality, which are called Service Level Objectives (SLOs). Latest advances in Edge computing help ensure SLOs because they extend the IoT domain with powerful, decentralized computing infrastructure~\cite{morichetta2023intent} to process IoT data with low latency. 
Each service in a larger smart city application has specific SLOs to fulfill, which help maintain the overall equilibrium.
This work focuses on a scenario where an Edge node provisions the execution of multiple services. We envision a smart city use case where connected devices, such as a video camera and a vehicle, each provide data to a dedicated service for workload execution at the Edge, as illustrated in Figure~\ref{fig:high-level-idea}.

% Cloud has a lot of resource, Edge does not
Transitioning from the Cloud to decentralized Edge computing platforms can improve real-time processing and privacy preservation; however, there is one thing that Edge computing cannot (yet) provide to the same extent: resources. Ensuring SLOs in the Cloud has a long history, where the successful recipe~\cite{verma_auto-scaling_2021} is to allocate additional resources in case the performance is mitigated. 
Edge devices can also be equipped with powerful hardware -- thus often called Edge servers -- but their ability to scale according to changing demands inevitably hits a resource limit. 
To circumvent this, it is possible to compose Edge and Cloud computing layers -- called the Computing Continuum (CC)~\cite{dustdar_distributed_2023_short} -- which can provide collaboration between devices, e.g., offloading load within the CC. This work, however, excludes offloading and focuses on methods for guaranteeing SLO fulfillment at a single Edge device.

Ensuring SLOs at the Edge requires alternative elasticity strategies -- apart from provisioning additional resources. To provide more flexibility, the solution can be to scale services within three \textit{dimensions}~\cite{dustdar_principles_2011}: quality, cost, and resources. 
In this sense, multi-dimensional scaling strategies~\cite{sedlak_controlling_2023_short,laso_multidimensional_2025} can dynamically decide how to optimize the application. For example, in case there are unclaimed resources, provide them to a service, or otherwise, scale down the quality; this is shown in Figure~\ref{fig:high-level-idea}.
Contrarily, existing autoscalers, even if dynamic, solely scale resources. This can combine vertical and horizontal scaling into a hybrid approach, as done by Lombardi et al.~\cite{lombardi_elastic_2018_short}, or adjusting SLOs based on environmental changes, e.g., when additional clients join, as done by Horovitz et al.~\cite{horovitz_efficient_2018_short}.
Lightweight scaling solutions for serverless functions, as done by Zhao et al.~\cite{zhao_tiny_2022}, also help to ensure SLOs for stream processing. Although these solutions do target resource-restricted devices, they do not harness the potential of other elasticity dimensions in scenarios where no more unoccupied resources can be provisioned.

To fill this gap, this work proposes a novel Edge-based autoscaler that elastically scales stream processing services in two dimensions: resources and quality. Our architecture consists of two layers of scaling agents: in the first layer, each service is extended with a Deep Reinforcement Learning (DRL) agent, which optimizes local SLO fulfillment by scaling resources or quality. These service-specific agents act greedily, which means that they can claim resources that other services might need. Hence, if the processing resources are exhausted, a higher-layer agent tries to improve global SLO fulfillment by reallocating resources.
Our evaluation underlines how this approach can improve SLO fulfillment under tight resource constraints, which motivates integrating it into existing CC platforms~\cite{nastic2022sfc}.

\section{Methodology}

In this section, we first describe the processing environment in which the scaling agents are embedded. Then, we present the design of the agents, focusing on how they interact with the environment for sensing, model training, and decision-making.

\subsection{Processing Environment}

%Some of the core entities have already been presented in Fig.~\ref{fig:high-level-idea}, namely 
Together, the data \textit{provider}, processing \textit{service}, and data \textit{consumer} (cfr. Fig.~\ref{fig:high-level-idea}) form a generalizable end-to-end streaming pipeline in which data streams from IoT devices are processed on nearby Edge devices and relayed to end users.

\textbf{Service definition} We define a \textit{service} as $s = \langle f, C, M, Q \rangle$, which describes the processing function ($f$), the service configuration ($C$), and metrics ($M$) generated during execution. The quality of a processing service, such as video resolution, is adjusted through the service configuration.
$Q$ contains a list of SLOs that should be fulfilled during runtime, where a single SLO $q \in Q$ is defined by $q = \langle v, rel, t, w \rangle$. This implies that a variable ($v$) should either be higher or lower ($rel$) than a threshold ($t$). For example, a video processing service could aim for a satisfactory frame rate through $q = \langle \textit{fps}, >, 25, 1.0 \rangle$, or save energy by limiting its allocated CPU cores through $q = \langle \textit{cores}, <, 5, 0.5 \rangle$ These SLOs are ranked according to a weight ($w$), which will be used by the scaling agents to rank its objectives. Given a metric $m \in M$ that reflects variable $v$, the SLO fulfillment ($\phi$) is calculated as in Eq.~\eqref{eq:slo-f}.
\begin{equation}
    \label{eq:slo-f}
    \phi(q, m) = 
    \begin{cases}
    \frac{m}{t},& \text{if}\  rel = '>' \\
    1 - \frac{m}{t},& \text{if}\  rel = '<'
    \end{cases}
\end{equation}
Metrics on a scale $m \in [0,\infty)$, such as \textit{fps}, thus produce values that start from 0.0 (= 0\% fulfillment) and can exceed 1.0. Values for \textit{cores}, however, fall into $m \in [1,c_{phy}]$ with $t=c_{phy}$. Thus, allocating fewer cores provides higher SLO fulfillment.

\textbf{Granular SLO fulfillment} Most established Cloud platforms~\cite{verma_auto-scaling_2021} employ binary logic to determine SLO fulfillment. In contrast, Eq.~\ref{eq:slo-f} provides a fuzzy ratio quantifying the extent of the SLO fulfillment. We favor this approach as it enables a more fine-granular elasticity control for autoscalers.

\textbf{Service execution} The service execution on an Edge device is wrapped in a Docker container.
% that can be executed in the runtime environment at a target device. 
% Notice, that most common Edge devices, such as the NVIDIA Jetson\footnote{www.nvidia.com/en-us/autonomous-machines/embedded-systems/}, support such containers by default, which has a central benefit: 
Thus, container resources can be vertically scaled by adjusting the number of allocated CPU cores. Notice, how Edge devices are constrained in numerous other ways (e.g., network or GPU), which can be scaled in future work. % it is important to consider the maximum amount of resources that Edge devices have integrated; while numerous resources are shared between services, this paper focuses on CPU allocation. 
In the context of this paper, we define a device $d = \langle c_{phy} \rangle$ through its number of physical CPU cores.

\subsection{Design of Scaling Agents}

%Now that it is clear how the processing environment is structured, 
Within this processing environment, we develop a two-layer scaling solution based on a hierarchical setting of agents that ensures processing SLOs on Edge devices through multi-dimensional elasticity; Fig.~\ref{fig:high-level-methodology} provides a high-level overview:
(1) Processing services are monitored to collect their states and respective SLO fulfillment; (2) each service is then extended with a Local Scaling Agent (LSA), which learns a scaling policy that optimizes SLO fulfillment; (3) if SLOs are violated, the LSA infers in which dimension to scale its service. When all resources are allocated, (4) resources can only be scaled through a mediator -- called Global Service Optimizer (GSO) -- which optimizes global SLO fulfillment by swapping resources.
%
% At the service level, we design a Local Scaling Agent (LSA) which trains a decision model over continuous services processing metrics to infer optimal scaling actions. If the agents behaves greedily, e.g.,occupying all the resources, a higher-level agent -- called Global Service Optimizer (GSO) -- improves global SLO fulfillment.
%
In the following, these steps are further elaborated:

\begin{figure}[t]
    \centering
    \includegraphics[width=1.0\linewidth]{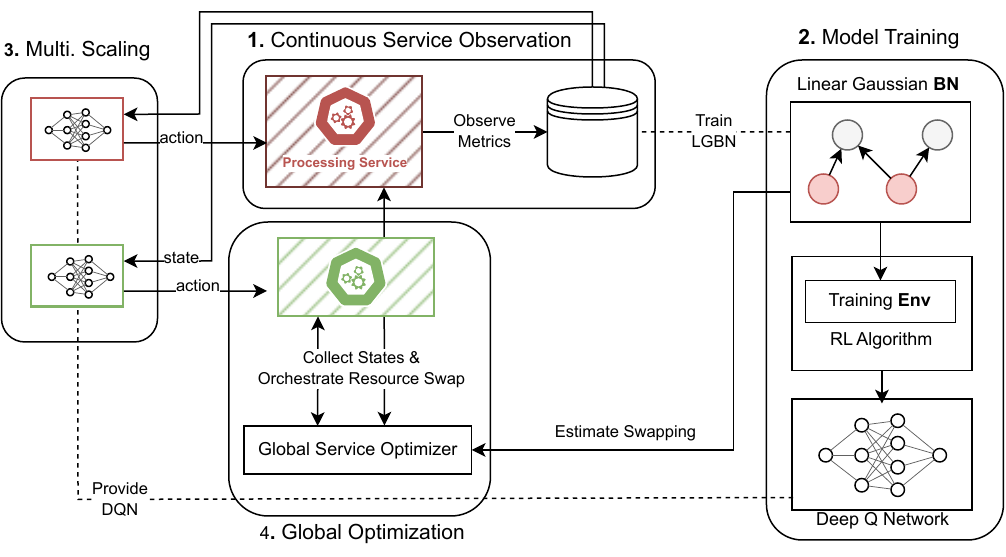}
    \caption{High-level view of the three-step methodology; continuously observing service executions, training an inference model, and using it for multi-dimensional scaling}
    \label{fig:high-level-methodology}
\vspace{-6pt}
\end{figure}

\subsubsection{Continuous Service Observation}

During runtime, every service periodically logs a snapshot of its state to a local buffer, which will later be collected by the LSA. Recall, that a service's state includes its configuration, metrics, and SLOs.
% hence, the agent does not need to repeatedly collect the metrics for persistence, but only access them when needed.
Depending on the use case, streaming services, e.g., video or audio processing, might log their state after processing one batch or frame. This creates a history of how a processing system was configured at a specific time, and to what degree ($\phi$) its SLOs were fulfilled.
% We will use this information in the next step to train the LSA through DRL. 
%
Since processing services and scaling agents (i.e., LSAs and GSO) are executed on the same physical devices, the metrics can be accessed by all of them.
% While this information provides detailed historical insight, scaling agents must act quickly based on the global device services, e.g., the resource allocation of all services. For this, the states of the services are also collected in a time-series DB.

% When scaling agents need to choose an action, having access to the latest state allows them to make accurate decisions.

\subsubsection{Model Training}

The LSA has one central objective: fulfill the SLOs ($Q$) of its assigned service. 
% To solve this multi-objective problem, the LSA must scale the right dimension at the right time. 
To find an optimal scaling policy for this multi-objective problem, the LSA is trained through Deep Reinforcement Learning (DRL), where it is rewarded for actions that lead to satisfying service states.
The sweet spot is to fulfill SLOs with $\phi_{opt} =1.0$, i.e., without overprovisioning resources or sacrificing too much quality. 
Hence, the agent aims to minimize the difference ($\Delta$) between $\phi$ and $\phi_{opt}$, which is expressed by Eq.~\ref{eq:slof_ll}, where $w$ is used to scale the reward of each SLO.
%
%\sum_{s \in S}
\begin{equation}
\label{eq:slof_ll}
    \Delta \gets  \sum_{q \in Q} | \phi_{opt} - \phi (q,m) | \times w_q
\end{equation}

% To learn how different scaling actions impact $\Delta$, the LSA learns a policy through DRL. However,

\textbf{Training Environment} Model-free RL algorithms often require tens of thousands of iterations to converge to a satisfying result -- this is not compatible with our processing environment because the effects of scaling actions, including rewards, are only reflected with a delay -- prolonging each training cycle considerably.
% also, (2) to stabilize learning for fluctuating performance, the service state must be evaluated in a sliding window.
To address this, we simulate state transitions and respective rewards through a virtual training environment\footnote{As described by OpenAI's \href{https://gymnasium.farama.org/introduction/create_custom_env/}{Gymnasium} environments}.
To create this environment, the LSAs use historical service metrics ($M$), however, cutting out periods of two seconds\footnote{We observed that roughly after this time scaling actions are reflected} after an action took place. The remaining metrics are used to train a Linear Gaussian Bayesian Network (LGBN) that expresses the relations between system variables.
% Note, that this builds on the assumption that the relations are linear. 
For a video processing task, the LGBN can express how processing $\textit{fps}$ depends on video resolution (\textit{pixel}) and provisioned \textit{cores}. Given these continuous variable relations, we can estimate the state transitions and rewards for hypothetical actions.

\textbf{Training Actions} The LSAs use this training environment to learn a scaling policy through a Deep Q Network (DQN), where the agent can take 5 different actions: (1) do nothing if its current state is satisfactory, (2$|$3) increase or decrease \textit{pixel} by $\pm \delta_{pixel}$, or (4$|$5) adjust \textit{cores} by $\pm \delta_{cores}$. The LSA should maintain these values in the range $\textit{pixel} \in [p_{min}, p_{max}]$ and $\textit{cores} \in [1,c_{free}]$, where $c_{free}$ is the number of unclaimed cores, with $c_{free} \leq c_{phy}$.
Notice, that $\delta_{pixel}$ and $\delta_{cores}$ are both constant factors, which we plan to substitute with continuous actions in future work.
The LSA uses these actions to interact with the training environment; as DRL converges, the DQN estimates the SLO improvement given an action and state. 
% If an action creates a state outside these ranges, this is punished. 

%
\textbf{Model Retraining} The LGBN is very dependent on the collected metrics, which is problematic because (1) no metrics are available at service start, and (2) historical metrics become outdated due to variable drifts. Hence, LGBN and DQN must be retrained periodically, which is done in the background with restricted resources. This retraining frequency can either be decreased gradually or coupled to SLO fulfillment.

\subsubsection{Multi-dimensional Scaling}

During runtime, the LSA uses its latest DQN to infer how to scale its attached service. Given the current service state, the DQN produces one of the five trained actions, which means that the LSA now interacts with the physical processing environment by either: scaling the service quality as part of the service configuration or scaling the resources by adjusting the container limitations.

\subsubsection{Global Optimization}

To improve the SLO fulfillment -- and minimize $\Delta$ -- the LSAs act greedy when they scale resources, which can be unfair to the other tenants on the Edge device. However, even if all resources have been allocated, it is possible to improve global (i.e., device-wide) SLO fulfillment. For instance, swapping a core from a service $a$ to a service $b$ that operates under tight SLO boundaries may improve the global SLO fulfillment because $a$'s gain is higher than $b$'s loss. Finding such operations is the responsibility of the GSO.

To decide whether it would be beneficial to swap a core between services, the GSO uses the available LGBNs. Given the states of both services ($a,b$), the GSO can again estimate what the expected state transition and consecutive reward would be when either swapping a core from $a \rightarrow b$, or $b \rightarrow a$. If the GSO estimates that one of these options improves global SLO fulfillment, it updates the container limits accordingly.

\section{Experimental Evaluation}

To evaluate the presented scaling solution, we create an instance of the processing environment and develop a physical prototype of our scaling agents that we apply during two experimental scenarios. Namely, we evaluated: (1) How the LSA performs under tight resource constraints compared to established autoscalers, and (2) how the GSO can optimize the global SLO fulfillment after all resources were allocated.

\subsection{Experimental Setup}
\label{subsec:setup}

To create a realistic stream processing scenario, we use OpenCV to continuously transform a video stream. The implementation of the processing service and the evaluation scenarios are publicly available on GitHub\footnote{\href{https://github.com/borissedlak/multiScaler}{github.com/borissedlak/multiScaler}}. In the repository, you also find the description of how the Computer Vision (\textit{CV}) service is containerized and how it is scaled during deployment.
% the evaluation scenarios will use multiple instances of \textit{CV}, but with different SLO thresholds to create heterogeneity.

Table~\ref{tab:slos} shows three types of SLOs ($Q$) for the \textit{CV} service: to guarantee the quality of experience, the LSA should ensure high video resolution (\textit{pixel}) and streaming framerate (\textit{fps}). To save energy, the LSA also minimizes the number of allocated \textit{cores}; however, this SLO can be traded off in favor of performance -- hence it has a lower weight ($w$). Notice, how \textit{fps} also has a higher weight than \textit{pixel}\footnote{During evaluation, these weights showed to reflect the intended tradeoff}. For each SLO $q\in Q$, the thresholds ($t$) are specified in the following scenario.

% \ins{missing start from random pixel/fps}

\setlength{\tabcolsep}{3.5pt}
\begin{table}[t]
    \small
  \centering
  \caption{Constrained variables for the \textit{CV} service}
  \label{tab:slos}
  \begin{tabular}{clccc}
    \toprule
    Var. & Description & Rel. & Weight & Impact \\
    \midrule
    \textit{pixel} & video streaming resolution & $> t$ & 0.8 & \textit{pixel} $\rightarrow$ \textit{fps} \\
    \textit{cores} & container usable CPU cores & $< 10$ & 0.4 & \textit{cores} $\rightarrow$ \textit{fps} \\
    \textit{fps} & processing throughput & $> t$ & 1.2 & -- \\
    \bottomrule
  \end{tabular}
  \vspace{-10pt}
\end{table}

% \tmi{What tools are applied in the architecture? I.e., Prometheus, Grafana, etc.}

% \tmi{What was the structure of the NN?}

% \tmi{How was the LGBN structure specified?}

\subsection{Service Scaling under Resource Constraints}

% Setup

This scenario starts with an inactive LSA that does not yet possess any models (i.e., LGBN or DQN). The LSA awaits 30s of \textit{CV} processing and then starts the first of five phases, in which: (1) the thresholds for \textit{fps} and \textit{pixel} are adjusted as shown in Tab.~\ref{tab:rounds}; to restrict resources, we also adjust $c_{phy}$. The LSA then (2) trains the LGBN and DQN from all existing metrics and (3) operates 50s in the environment by autoscaling the \textit{CV} service. This concludes the first phase; the next 4 phases are conducted the same way. To increase the stability of our results we repeat this entire scenario 5 times.
We compare the LSA's effective SLO fulfillment with a baseline, similar to the Kubernetes VPA: initially, the VPA assigns $\textit{pixel} = t$, which means it cannot sacrifice service quality. During runtime, the baseline VPA scales resources according to $\phi(fps)$, i.e., scaling $cores + 1 $ if $\phi(fps) < 1.0$ of $cores -1$ if $\phi(fps) > 1.0$
%
% To increase the stability of our experiment we repeat this scenario 5 times.

% Result

\begin{figure}[t]
    \vspace{-5pt}
    \centering
    \includegraphics[width=1.0\linewidth]{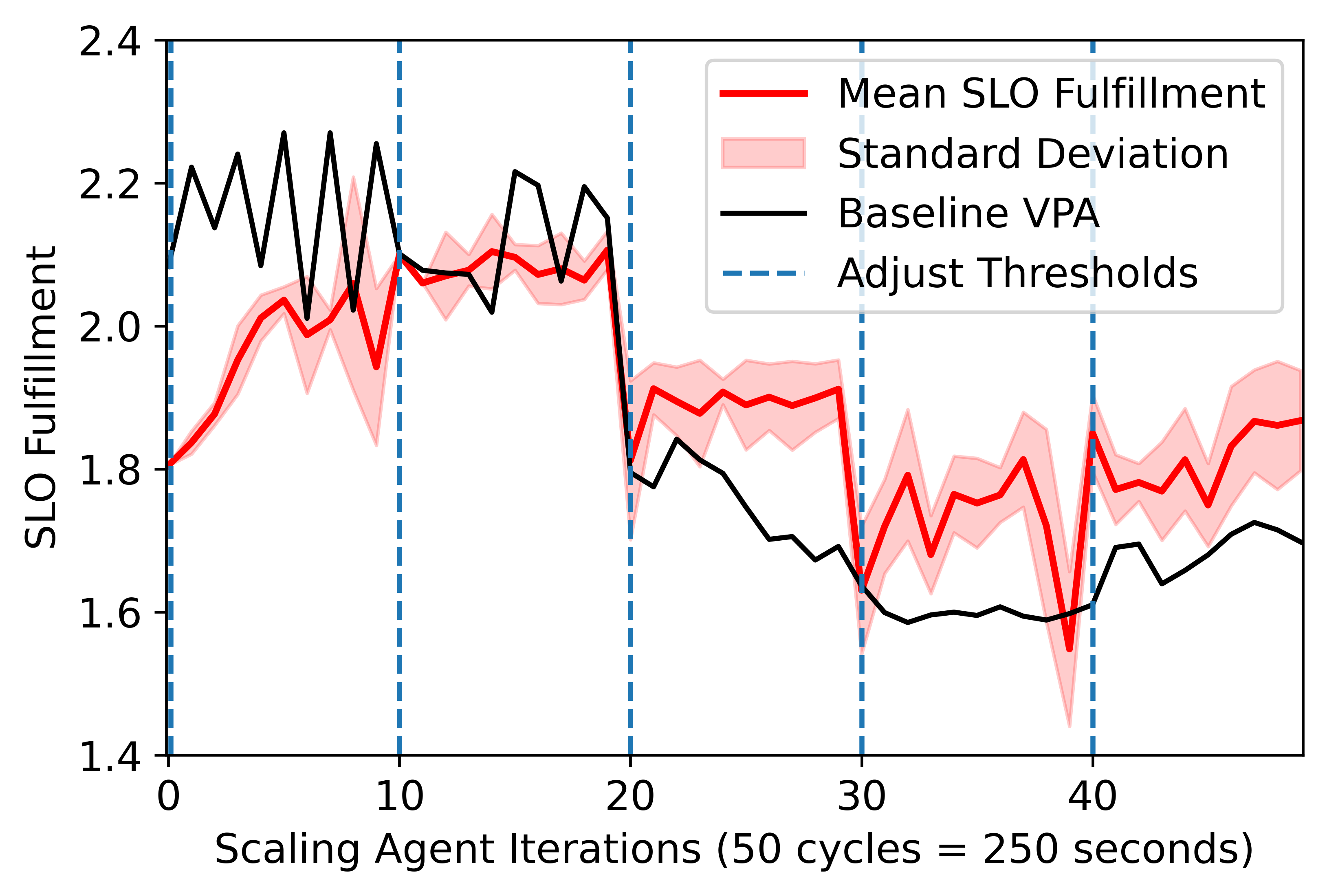}
    \caption{SLO Fulfillment during runtime; every 10 iterations the SLO thresholds and available resources are changed} 
    \label{fig:tight-constraints}
    \vspace{-8pt}
\end{figure}

As depicted in Fig.~\ref{fig:tight-constraints}, the LSA initially performed slightly under the baseline, when its DQN was not yet accurate in the first two phases. In the subsequent rounds, however, the LSA outperformed the baseline VPA because it was able to trade off parts of the \textit{pixel} SLO to fulfill the higher-weighted \textit{fps} SLO. Notice, how the y-axis shows the cumulative SLO fulfillment ($\phi_{\Sigma}$) with a maximum of $\phi_{\Sigma} \leq \sum_{q \in Q} w$; hence $\leq 2.4$

Due to page limitations, we cannot provide detailed results on the training overhead, still, we would like to comment that we limited the training to one core. Within these limited resources, LGBN training took roughly 1s, and roughly 10s for the DQN. Suppose we train the models less frequently than every 50s the overhead should not impact performance.

\setlength{\tabcolsep}{4pt}
\begin{table}[t]
    \small
  \centering
  \caption{SLOs as variable constraints for the \textit{CV} service}
  \label{tab:rounds}
  \begin{tabular}{crrrrr}
    \toprule
    Var. & Phase 1 & Phase 2 & Phase 3 & Phase 4 & Phase 5\\
    \midrule
    \textit{pixel} & $> 800$ & $> 1000$ & $> 1700$ & $> 1900$ & $> 1800$\\
    \textit{fps} & $> 33$ & $> 33$ & $> 35$ & $> 35$ & $> 34$\\
    \textit{max core} & 9 & 7 & 8 & 2 & 3\\
    \bottomrule
  \end{tabular}
\end{table}

% Implications

\implication{With sufficient training, the LSA can be applied in resource restricted scenarios to choose between multi-dimensional elasticity strategies; this showed to improve SLO fulfillment further than a regular VPA}

\subsection{Global Optimization after Resource Exhaustion}

% Setup

The following scenario evaluates if the GSO can optimize global SLO fulfillment (= $\phi_{\Sigma,Alice} + \phi_{\Sigma,Bob}$) when no free resources can be allocated: We start two instances of CV -- called Alice and Bob  -- which are supervised by two LSAs. Both LSA should ensure an SLO for $\textit{pixel} > 1300$; additionally, we put a tight SLOs for Alice with $\textit{fps} > 30$, whereas Bob only requires $\textit{fps} > 10$. As soon as all resources are exhausted, the GSO takes action. Notice, how $\Delta = (\sum_{q \in Q} w) - \phi_{\Sigma}$.

% Result

\begin{figure}[t]
    \vspace{-8pt}
    \centering
    \includegraphics[width=1.0\linewidth]{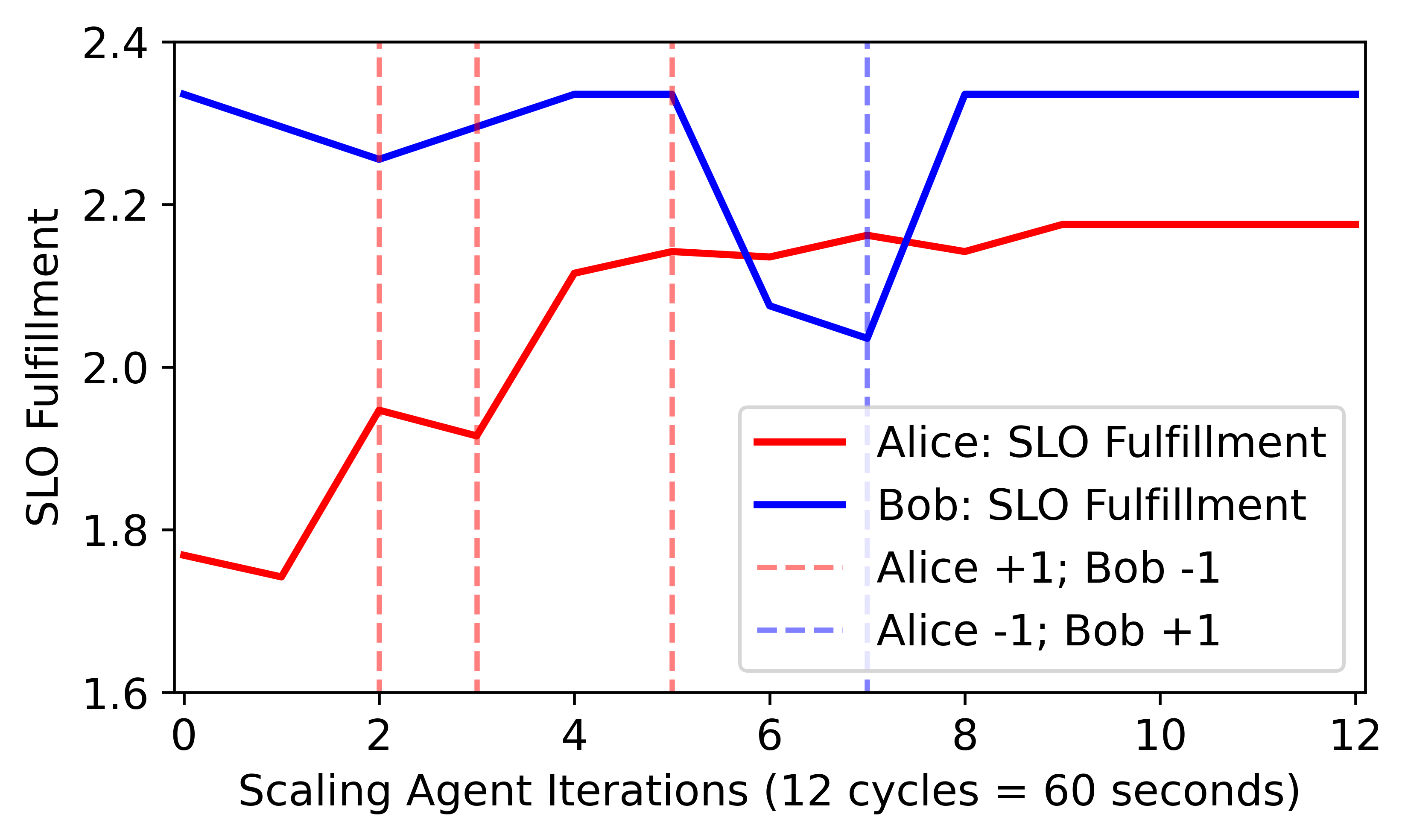}
    \caption{SLO fulfillment of two services operating with resource contention; the GSO swaps resources to globally improve SLOs}
    \label{fig:global-optimizer}
    \vspace{-6pt}
\end{figure}

As depicted in Fig~\ref{fig:global-optimizer}, the GSO decides in iterations $i=2$ and $i=3$ to swap a core from Bob $\rightarrow$ Alice, which improves $\phi_{\Sigma,Bob}$, while showing no notable impact on $\phi_{\Sigma,Alice}$. Thus, increasing global SLO fulfillment. However, at $i=5$, the same operation did not provide the expected result because it would harm Alice and not provide much benefit for Bob. This blunder, however, is resolved shortly after, when at $i=7$ the GSO decided to swap back a core from Alice $\rightarrow$ Bob.

% Implication

\implication{In situations where no unclaimed resources are available, the GSO can improve the global SLO fulfillment by shifting resources between greedily acting autoscalers}

% \section{Discussion}

% \ins{Either keep this a full section or incorporate into Evaluation}

\section{Conclusion}
In this paper, we presented a multi-dimensional scaling solution that ensures processing SLOs on Edge devices. To improve SLO fulfillment in resource-constrained devices, we built a two-layer architecture with local, service-specific agents that scale either quality or resources; if all resources are allocated, a global mediator swaps resources between services to further improve SLO fulfillment. The innovation of our solution relies on the injection of domain-specific knowledge, i.e., the relation between system variables through an LGBN, and scaling processing services in multiple elasticity dimensions depending on the context. For future work, we have a clear research agenda for which we plan to improve the following aspects: (1) eliminate the DQN to operate directly on the LGBN to infer scaling actions, (2) produce continuous scaling actions for fine-grained control, (3) extend the architecture with more edge devices to support offloading, and (4) use scenarios with different, heterogeneous stream processing services.

\section*{Acknowledgement}

Funded by the European Union (TEADAL, 101070186). 
%Views and opinions expressed are those of the authors only and do not necessarily reflect those of the EU. Neither the EU nor the granting authority can be held responsible for them.

\bibliographystyle{IEEEtran}
\bibliography{Boris,Andrea}

\begin{thebibliography}{10}
\providecommand{\url}[1]{#1}
\csname url@samestyle\endcsname
\providecommand{\newblock}{\relax}
\providecommand{\bibinfo}[2]{#2}
\providecommand{\BIBentrySTDinterwordspacing}{\spaceskip=0pt\relax}
\providecommand{\BIBentryALTinterwordstretchfactor}{4}
\providecommand{\BIBentryALTinterwordspacing}{\spaceskip=\fontdimen2\font plus
\BIBentryALTinterwordstretchfactor\fontdimen3\font minus \fontdimen4\font\relax}
\providecommand{\BIBforeignlanguage}[2]{{%
\expandafter\ifx\csname l@#1\endcsname\relax
\typeout{** WARNING: IEEEtran.bst: No hyphenation pattern has been}%
\typeout{** loaded for the language `#1'. Using the pattern for}%
\typeout{** the default language instead.}%
\else
\language=\csname l@#1\endcsname
\fi
#2}}
\providecommand{\BIBdecl}{\relax}
\BIBdecl

\bibitem{chui2021internet}
M.~Chui, M.~Collins, and M.~Patel, ``The internet of things: Catching up to an accelerating opportunity,'' 2021.

\bibitem{morichetta2023intent}
A.~Morichetta, N.~Spring, P.~Raith, and S.~Dustdar, ``Intent-based management for the distributed computing continuum,'' in \emph{IEEE SOSE}, 2023.

\bibitem{verma_auto-scaling_2021}
S.~Verma and A.~Bala, ``\BIBforeignlanguage{en}{Auto-scaling techniques for {IoT}-based cloud applications: a review},'' \emph{\BIBforeignlanguage{en}{Cluster Computing}}, vol.~24, no.~3, Sep. 2021.

\bibitem{dustdar_distributed_2023_short}
S.~Dustdar, V.~C. Pujol, and P.~K. Donta, ``On {Distributed} {Computing} {Continuum} {Systems},'' \emph{IEEE TKDE}, Apr. 2023.

\bibitem{dustdar_principles_2011}
S.~Dustdar, Y.~Guo, B.~Satzger, and H.-L. Truong, ``Principles of {Elastic} {Processes},'' \emph{Internet Computing, IEEE}, vol.~15, pp. 66--71, Nov. 2011.

\bibitem{sedlak_controlling_2023_short}
B.~Sedlak, V.~Casamayor~Pujol, P.~K. Donta, and S.~Dustdar, ``Controlling {Data} {Gravity} and {Data} {Friction}: {From} {Metrics} to {Multidimensional} {Elasticity} {Strategies},'' in \emph{2023 {IEEE} {Services}}, Jul. 2023.

\bibitem{laso_multidimensional_2025}
S.~Laso, I.~Murturi, P.~Frangoudis, J.~L. Herrera, J.~M. Murillo, and S.~Dustdar, ``A {Multidimensional} {Elasticity} {Framework} for {Adaptive} {Data} {Analytics} {Management} in the {Computing} {Continuum},'' Jan. 2025.

\bibitem{lombardi_elastic_2018_short}
F.~Lombardi, L.~Aniello, S.~Bonomi, and L.~Querzoni, ``Elastic {Symbiotic} {Scaling} of {Operators} and {Resources} in {Stream} {Processing} {Systems},'' 2018.

\bibitem{horovitz_efficient_2018_short}
S.~Horovitz and Y.~Arian, ``Efficient {Cloud} {Auto}-{Scaling} with {SLA} {Objective} {Using} {Q}-{Learning},'' in \emph{2018 {FiCloud}}, Aug. 2018.

\bibitem{zhao_tiny_2022}
Y.~Zhao and A.~Uta, ``\BIBforeignlanguage{English}{Tiny {Autoscalers} for {Tiny} {Workloads}: {Dynamic} {CPU} {Allocation} for {Serverless} {Functions}}.''\hskip 1em plus 0.5em minus 0.4em\relax IEEE Comp. Society, 2022.

\bibitem{nastic2022sfc}
S.~Nastic, P.~Raith, A.~Furutanpey, T.~Pusztai, and S.~Dustdar, ``A serverless computing fabric for edge and cloud,'' in \emph{IEEE CogMI}, 2022.

\end{thebibliography}

\end{document}